# Laser-patterned multifunctional sensor array with graphene nanosheets as a smart biomonitoring fashion accessory

*Vijay Shirhatti[1], Suresh Nuthalapati[1], Vaishakh Kedambaimoole[1], Saurabh Kumar[2], Mangalore Manjunatha Nayak[2], Konandur Rajanna[1]\**

[1]Dept of Instrumentation and Applied Physics, Indian Institute of Science, Bangalore 560012, India

[2]Centre for Nano Science and Engineering, Indian Institute of Science, Bangalore 560012, India

Corresponding Author: K Rajanna, Email: kraj@iisc.ac.in

**Abstract**

Biomonitoring wearable sensors based on two-dimensional nanomaterials have lately elicited keen research interest and potential for a new range of flexible nanoelectronic devices. Practical nanomaterial-based devices suited for real-world service, which have first-rate performance while being an attractive accessory, are still distant. We report a multifunctional flexible wearable sensor fabricated using an ultra-thin percolative layer of microwave exfoliated graphene nanosheets on laser-patterned gold circular inter-digitated electrodes for monitoring vital human physiological parameters. This Graphene on Laser-patterned Electrodes (GLE) sensor displays an ultra-high strain resolution of $245\mu\varepsilon$ (0.024%) and a record gauge factor of 6.3e7 and exceptional stability and repeatability in its operating range. The sensor was subjected to biomonitoring experiments like measurement of heart rate, breathing rate, body temperature, and hydration level, which are vital health parameters, especially considering the current pandemic scenario. The sensor also served in applications such as a pedometer, limb movement tracking, and control switch for human interaction. The innovative laser-etch process used to pattern gold thin-film electrodes and shapes, with the multifunctional incognizable graphene layer, provides a technique for integrating multiple sensors in a wearable fashion accessory. The reported work marks a giant leap from the conventional banal devices to a highly marketable multifunctional sensor array as a biomonitoring fashion accessory.

**Introduction**

        The human body exhibits ample cues regarding its present state of wellbeing. While the human body possesses the sensory and somatosensory perception of the outside world and its conditions, external devices are required to know the body's status accurately. Up until the turn of this century, a wearable device would just mean a wristwatch or a medical device in the intensive care unit. The swift progression in semiconductor and telecommunication technology has led to a new meaning to the term wearable devices ranging from wrist bands, arms bands, and so forth. Over the last decade, flexible wearable devices have emerged as the primary personal in-house health monitoring devices, which provide clues to the bearer key physiological parameters and help in timely medical intervention. The silicon technology does not cater to these applications principally due to the sensor's rigid structure and inability to comply with the human body shape. Monitoring of human physiological parameters still poses an uphill task and demands sensors with flexibility, stretchability and bio-compatibility, presently in a continual state of improvement.

        The most popular sensor employed in commercially available wearable devices is the Photoplethysmography (PPG) sensor used to measure heart rate and respiration rate,[1] which has been researched and developed for decades. The PPG yet encounters certain limitations such as getting affected by optical noise, skin tone dependency, lower sensitivity in low perfusion areas like arms, legs (sensitive on fingers only), longer settling time, delivery of moving average of heart rate, lack of intra-peak details, high power consumption and larger form factor.[2] Sensors dedicated to other physiological measurements such as thermometers and skin-impedance meters used for measurement of body temperature and dehydration, respectively, are still independent devices and have not been integrated into commercially available wearable devices.

        Newer sensors termed as e-skins, epidermal electronics, smart textiles are being fervently researched in the field of wearable devices. Sensors based on nanomaterials like graphene, $MoS_2$, MXene have emerged as exciting prospects in accomplishing better detection of physiological parameters.[3] These 2D materials offer considerable improvements in response time, sensitivity, skin conformity, power consumption, and so on, which are reckoning for a new generation of wearable devices. Graphene being the front-runner among 2D nanomaterials, extensive wearable device research has been accomplished based on this material. In its various forms, graphene has been used to develop Flexible wearable devices for

motion monitoring and biomedical applications using CVD grown graphene, [4–8] graphene skin-confirming electrodes, [9] capacitive strain sensors,[10–13] reduced graphene oxide,[14–17] and nanocomposites based on graphene and metals/semiconductors and polymers.[18–35] have been extensively reported in the recent past. Similarly, other functional materials like MXene,[36] gold nanoparticles,[37,38] AgNW,[39] PEDOT:PSS[40] and so forth, have also been employed in a variety of designs as wearable sensors for different biomedical applications. Devices based on nanomaterials have shown immense promise and have effected commendable progress in wearable device technology.

However, certain limitations in the works reported above can be pinned down for further improvement to make the wearable device technology competent and market-ready. The device sensitivity, an outcome of the sensor thickness, can be further improved in the low strain (<0.1%) regime.[3] Controlling the sensor layer thickness has always been a significant cause of concern since lower thickness leads to film cracking, unreliable electrical contacts, and mechanical instability. A uniform conductive sensing film prescribes higher film thickness (>μm) which compromises the sensor sensitivity and makes it opaque and stodgy. The counter-intuitive idea of using 2D nanomaterial in a 3D model can lead to the stacking of graphene sheets and hence squanders the benefits of a 2D material. Sensors that are ultra-thin and transparent such as those using CVD grown monolayer graphene, suffer from lower sensitivity, complex handling process, low scalability, limited customizability, and higher production cost. The lower sensitivity demands stronger adhesion of the sensor/electrodes with the skin surface, like double-sided adhesive tape,[6] tattoo adhesion.[36] or on-skin printing[41] for maximum strain transfer / electrical interfacing. This arrangement renders the sensor as a single-use device since removal of the strongly bonded sensor without irreparable damage is an improbable task.

The demonstrated sensors have seldom been tested as multifunctional devices, which can greatly enhance their versatility and valuation. The reported sensors, especially foam or sponge-structured, are inherently bulky, opaque, low skin conformal and unappealing. Sensors that have been tested for different physiological and tactile applications have fared satisfactorily for the individual tasks, but consolidating multiple sensors in an ensemble has not been addressed before. Fabricating an ensemble that is practical and fashionable enough to become a marketable accessory still needs to be solicited. The development of a highly sensitive, simple, scalable, skin-conformal, multifunctional, customizable, daily-wear integrated wearable device for biomonitoring and consumer applications is highly demanding and presented in the following work.

In the present work, we have designed, developed and demonstrated a multifunctional sensor using solution-processed graphene nanosheets to monitor human physiological parameters. A unique way to fabricate the sensor electrodes using a laser-etch process has been developed, which forms the essence and the aesthetics of the device. The graphene-ink drop-casted on gold electrodes fabricated using a unique laser-patterning process forms the essence and the aesthetics of the wearable sensor. The 2D graphene nanosheets form a conducting percolative network across a micro-channel created by laser etching of the gold thin film. The sensor's remarkable characteristics and demonstration toward measurements of physiological parameters like heart rate, breathing rate, limb movements, touch input, body temperature and dehydration have been reported. An aesthetically appealing multi-sensor ensemble based on the innovated technique has been presented as a trendy wearable bracelet.

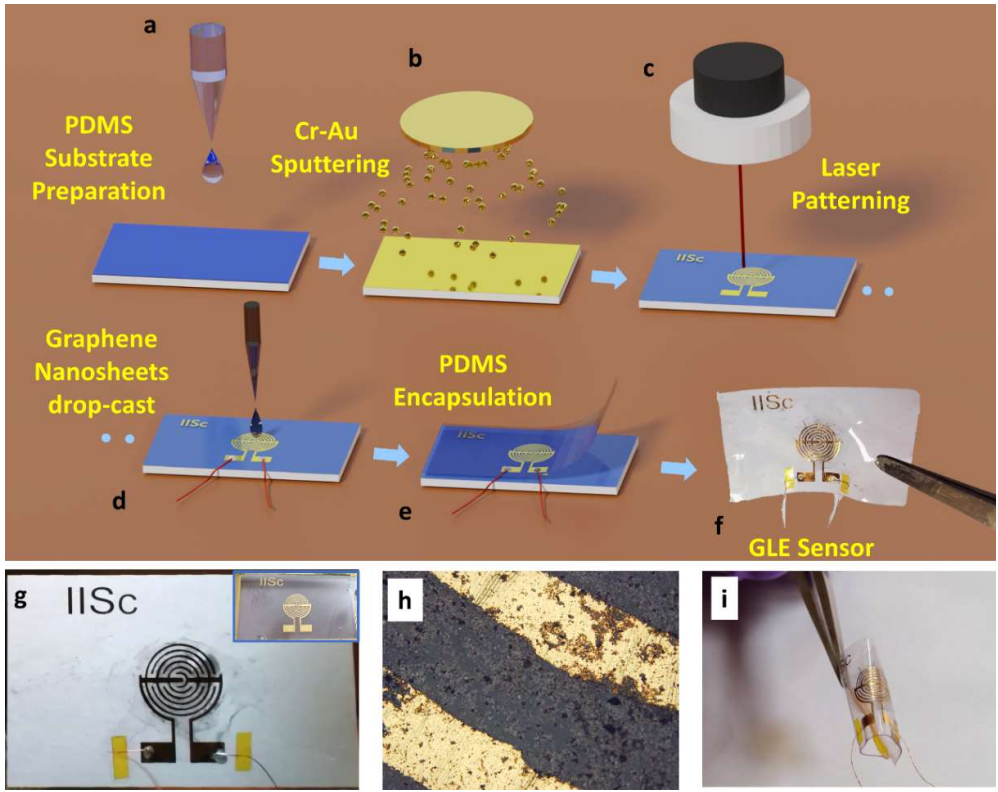

**Figure 1.** Schematic representation of fabrication process of single GLE sensor, (a) PDMS spin-coated on the glass substrate, (b) Deposition of Cr-Au thin-film via RF sputtering process, (c) Laser-patterning of the top thin-film layer to form IDE, exposing PDMS under-layer, (d) drawing electrical contacts and drop-casting GLE ink on the IDE, (e) encapsulation with top PDMS layer, (f) actual photograph of GLE sensor peeled from the glass substrate, (g) image of GLE sensor showing translucent graphene layer (inset) image of bare IDE, (h) magnified image of GLE sensor showing GLE sheets on IDE, (i) flexibility of GLE sensor.

**Results and Discussion**

**Material Characterization**

The physical and chemical properties of the as-synthesized MEGO material were ascertained via different material characterization techniques. The morphology of the MEGO material was examined under the FE-SEM tool, and an exploded form of GO was observed, much like an accordion instrument **(figure 2a)**. This structure is caused due to the intense agitation of the trapped functional groups on the surface of the carbon basal plane, during the graphene oxide synthesis process. The graphene oxide is rendered hydrophilic due to the oxidation process and hence readily adsorbs moisture on the surface. The microwave energy vibrates the functional groups and absorbed water molecules, which eventually split the loosely bonded graphene layers apart and escape. This mechanism leads to the accordion structured MEGO. The high magnification SEM image shows few-layer graphene (<3nm) separated from the stack **(figure 2b)**. To further decipher the graphene layers, Atomic Force Microscopy (AFM) was used. The AFM image reveals the presence of monolayer graphene nanosheets in the dispersion of MEGO **(figure 2c)**. The monolayer graphene nanosheet thickness is ~0.8nm, agreeing to available reports on the thickness of derived graphene nanosheets.[42] The thickness of the sensing layer on the IDE is approximated to ~10nm, extending up to 100 nm for few stacked or un-exfoliated particles **(figure S3)**.

The graphene nanosheets were examined under Tunneling Electron Microscope (TEM) and wrinkly graphene sheets were noticed **(figure 2d, 2e)**. The interlayer spacing was found to be around 0.39nm and the SAED pattern revealed the 6-fold pattern associated to the hexagonal lattice characteristic of the graphene crystalline structure.

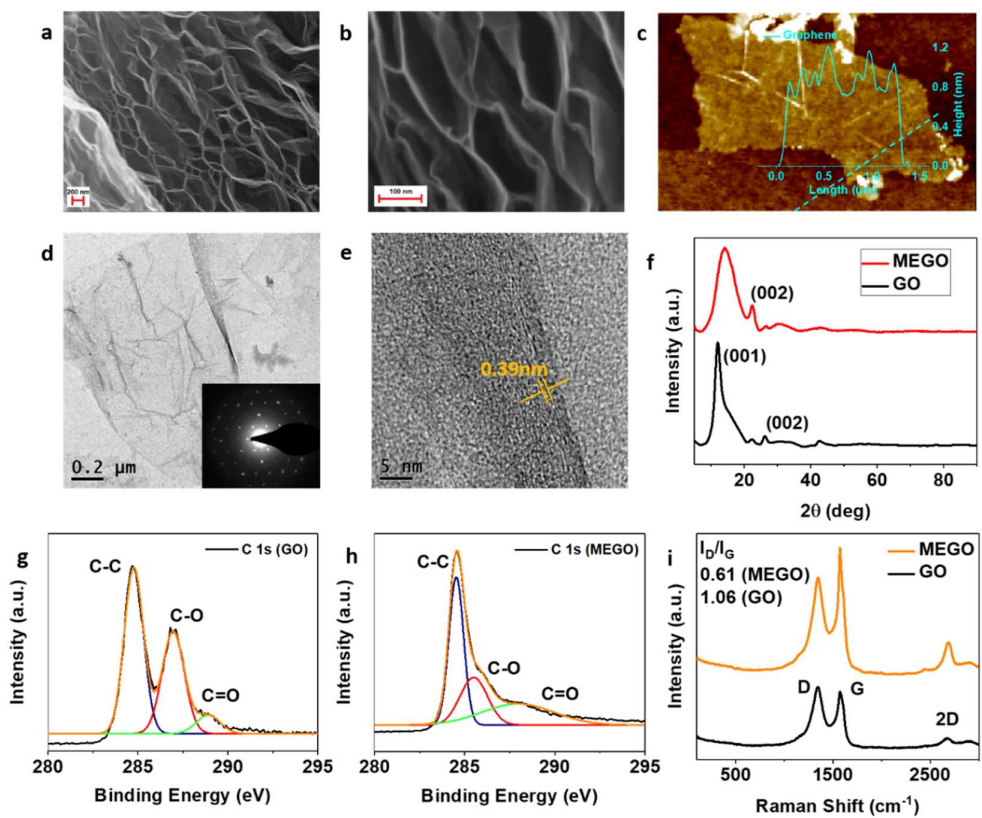

Figure 2. (a) FE-SEM image of MEGO showing the accordion-like structure, (b) high magnification FE-SEM image showing separated graphene sheets with <3nm thickness, (c) AFM image of monolayer graphene sheet with a mean height of ~0.8 nm and ~3μm lateral dimensions, (d),(e) TEM image of graphene sheets, (inset) graphene SAED pattern, and graphene sheet thickness measured at ~0.39 nm. (f) XRD analysis of GO and MEGO indicating (001) and (002) peaks, revealing reduction of GO after microwave thermal treatment, (g),(h) XPS analysis of GO showing 1Cs peak comprising C-C, C-O and C=O bonds, MEGO showing 1Cs peak with lower C-O and C=O peak intensity due to reduction of GO, (i) Raman spectroscopy of GO and MEGO indicating a reduction in $I_d/I_g$ ratio from GO to MEGO due to removal of functional groups and restoration of defects in the graphene basal plane due to microwave treatment.

The crystalline structure and the interlayer spacing of the synthesized materials was analyzed via X-Ray Diffraction (XRD) technique. The XRD pattern of GO shows a prominent peak near 11.9° corresponding to the (001) plane and an interlayer spacing of 0.73nm **(figure 2f and table S1)**. The XRD scan of MEGO revealed a broad peak ((001) plane) at 14.15° indicating reduction of GO and removal of the intercalated functional groups. A peak at 22.37° is observed corresponding to the (002) plane with a higher inter-layer spacing (0.4 nm) than graphite (0.34nm), peculiar of reduced graphene oxide as a result of residual functional groups on the graphene plane.[43]

The X-ray Photo Spectrometry (XPS) measurements definitively confirmed the reduction of GO due to the microwave treatment **(figure 2g, 2h)**. The C1s high-resolution scan of GO indicated prominent C-O and C=O bonds represented by peaks at 286.93 eV and 288.92 eV, respectively along with the dominant C-C peak at 284.6 eV. On the other hand, MEGO C1s high-resolution scan revealed reduced peak intensity of C-O and C=O bonds leading to the above inference.

The vibrational modes of the MEGO were observed in the Raman spectroscopy**(figure 2i)**. It was deduced that the removal of functional groups from the basal graphene plane due to microwave treatment resulted in the restoration of the defect sites. The $I_D/I_G$ ratio in MEGO was lower than that in GO, which led to the above conclusion.

The peculiar morphology of MEGO also warranted characterizing the material for its total surface area. The MEGO material was evidently higher in volume than the GO, and extremely light in weight such that the a mild air draft could blow its particles. The surface area of the MEGO was examined via N2 gas absorption and BET analysis. The material was found to have a total surface area of 351.1m$^2$/g **(figure S3)**.

**Mechanical & Thermal characterization**

The GLE sensor was subjected to known deformation using a Newport Micro-motion system. The sensor was fixed at the ends and deflected at the centre, i.e. at the sensor location, with known distances. A monotonic loading cycle shown in **figure 3b** displays increasing sensor resistance with the applied strains. The strain induced in the sensor was deduced from COMSOL Multiphysics modelling and simulation. A deflection of 0.5 mm in the sensor induces a strain of 2450 με. The induced strain for the applied sensor deflection was derived from COMSOL simulations (**figure 3a** and supplementary **figure S5**). The sensor's response in the form of change in resistance with respect to the mechanical inputs is recorded and analyzed **(figure 3c)**.

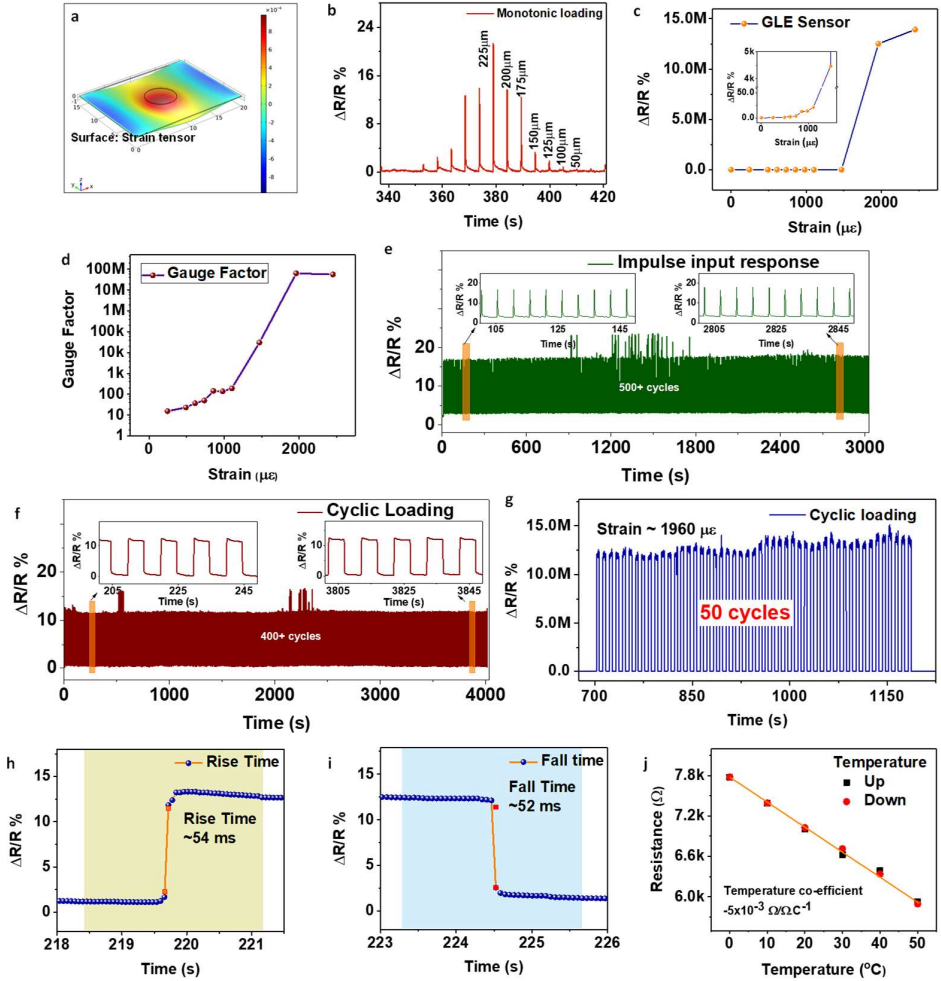

**Figure 3.** (a) COMSOL Multiphysics simulation for deriving strain induced in GLE sensor for known deflections, Strain distribution on the sensor surface, force applied at the sensor centre over the IDE, (b) Response of GLE sensor for monotonic sensor deflection, (c) exponentially increasing response of GLE sensor for applied strain, (d) plot showing sensor gauge factor variation with respect to applied strain, (e) Plots showing repeatability of GLE sensor for cyclic loading– Impulse/tapping mechanical input, (f) square-wave mechanical input, (g) Steady response of the sensor to multiple cycles of strain at high gauge factor(~6.3e7) conditions, (h) GLE sensor transient response- rise time (54 ms) and (i) fall time (52 ms), (j) temperature response of GLE sensor from 10 ºC to 50 ºC, calculated TCR ~ -0.005 $\Omega/\Omega.C^{-1}$.

The sensor had an ultra-high strain resolution of 245 µε (~0.024%), indicating immense potential in detecting delicate pressure signals. The sensitivity of a strain sensor is given in terms of Gauge factor (GF), i.e. ratio of change in resistance to applied strain (GF = (ΔR/R)/ε)

**(figure 3d)**. A GF of ~15 was observed at the above low strain input. This ultra-low detection limit was found to be better than most sensors reported in the literature so far. High sensitivity sensors, often described as feather touch-sensitive, have reported resolution of not better than 0.1% strain.[44–46] As the sensor is further strained, the resistance increases exponentially, and at ~1500 με, a 2000-fold increase in the GF is observed. Further increment in the applied strain results in a drastic increase in the sensor resistance reaching the megaohms range, resulting in a GF of 6.38e7 at ~2000 με. This magnitude of GF at the given strain input is unseen and exceeds reported values.[45,37,47,48]. The repeatability of the response was verified by subjecting it to more than 500 cycles of pulse-type mechanical input, as shown in **figure 3e,** and the response was found steady throughout the experiment. A consistent response was seen for square type cyclic loading experiment **(figure 3f)**. Repeatability of the sensor at higher strain ranges and higher GF has been an area of uncertainty. The conventional judgement indicates that sensor response with such high GF would be unstable, as reported in the literature,[46] where the sensitivity drastically reduces after a few cycles. To examine our sensor for such a behaviour, repeatability tests were carried out to inspect the sensor response at higher strain levels (~1950 με), where a gauge factor of ~6e7 was observed **(figure 3g)**. In our case, the sensor performed consistently for the cyclic loading experiments at different high-load conditions **(figure 3g, S6)**. These experiments adequately substantiate the repeatability and reliability of the sensor. The unprecedented GF and the consistency endorse the use of graphene in the sensor and the merit of the innovative design. The transient response of the GLE sensor was studied by inducing rapid deflection **(figure 3h, 3i)**. The sensor was deflected for 0.2 mm at a speed of 5 mm/s. The sensor's rise time was found to be 54 ms for an input rise time of 40ms, and the fall time of the sensor was found to be 52 ms for a similar input fall time.

The contribution of gold thin-film in the resistance change due to straining was also determined by carrying out experiments on the gold layer exclusively. The gold layer was strained till 5000με, and the corresponding change in resistance was recorded **(figure S7)**. The gauge factor was found to be ~2, which is peculiar of metal thin film type strain sensors. The thin film stayed intact for the magnitude of strain-induced during this experiment. It is, therefore, concluded that the GLE sensor performance is attributed entirely to the graphene nanosheets percolative networks and not to the circular gold IDE.

The GLE sensor fabricated on PET substrate and coated with PDMS layer was subjected to controlled temperature cycling. The temperature response of the sensor is shown in **figure 3j**. The response was linear for the range of concern, i.e. around the average human body temperature of 37°C. The sensor displayed negative temperature dependency with a temperature coefficient of resistance ~$5\times10^{-3}$ $\Omega/\Omega.C^{-1}$. Adopting the same graphene nanosheet material for temperature sensing application offers added advantage in ease-of-fabrication and a measure of temperature compensation for the strain sensors.

**Sensing mechanism:** The sensing mechanism of the device can be explained based on two principles; change in the contact area of the graphene nanosheets and the tunnelling mechanism between them. The deposition of graphene nanosheets on the sensor surface results in an arrangement where the nanosheets form a percolative network, bridging the IDE structure. Several such conductive paths are established by the nanosheets between the electrodes at different regions of the IDE. The graphene nanosheets are connected via weak van der Waal's forces and overlap each other at the edges forming ohmic contacts. The exerted strain forces the overlapped nanosheets to slide apart, thus reducing the sheets' aggregate contact area. This mechanism leads to a decrease in the conductivity of the percolative network giving a measurable manifestation of the induced strain.

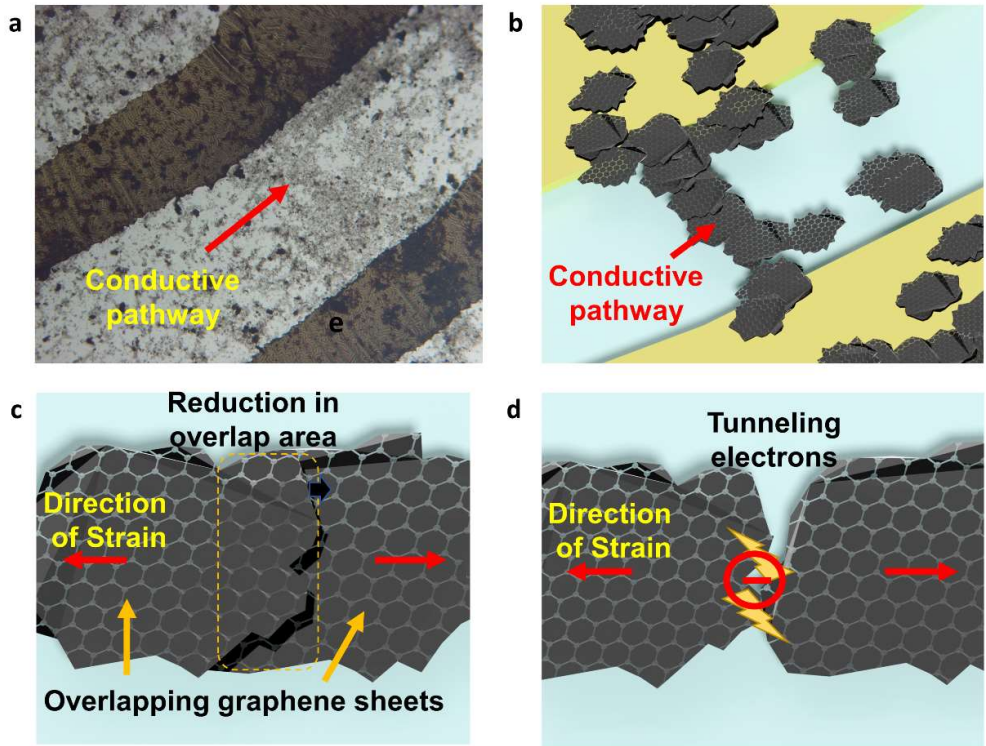

**Figure 4.** (a) Magnified backlit image of GLE sensor, percolative conductive pathway established between digits of the electrodes, (b) schematic showing percolative conduction pathway between the electrodes, (c) overlapped graphene nanosheets sliding apart due to applied strain, (d) tunneling phenomenon between the two graphene nanosheets when they are no longer in ohmic contact.

The mechanism can be graphically represented, as shown in **figure 4b, 4c** and mathematically represented as proportionality equation (1).

$$\Delta R \propto \frac{G}{P} \cdot \frac{1}{A_c}$$

…1

where $R$ is the resistance of the sensing film,

G is the number of graphene nanosheets forming a single conduction path,
P is the number of parallel conductive paths established between the two electrodes,
$A_c$ is the cumulative contact area between the graphene nanosheets in the network.

As the strain increases, the nanosheets slide further away, eventually reaching a point of no overlap but remain close enough (<3nm) to maintain tunnel contact. At this juncture, electrons tunnel through the polymer dielectric or void between the nanosheets, and the tunnelling mechanism ensues. A significant increase in the network resistance is witnessed with the rising strain induced in the material. The mechanism can be represented as shown in **figure 4d** and mathematically explained as per Simmons theory and model explained by Zhang et al.[49,50]

The resistance of the sensing film can be expressed as equation 2, shown below.

$$R = \frac{G}{P}\left[\frac{8\pi h d}{3 A_t \gamma e^2} \cdot exp(\gamma d)\right]$$

…2

where, G and P hold the same notations as above,

$h$ is the Planc's constant,

e is the electron charge,

$A_t$ is the effective area concerning the occurrence of tunnelling,

$\gamma = \frac{4\pi}{h}(2m\varphi)^{1/2}$ , where $m$ is the electron mass and $\varphi$ is the potential barrier between the two conducting particles, d is the distance between the adjacent graphene sheets.

The above equation can be simplified to the proportionality equation 3, since all other terms are approximated as constants,

$$R \propto \frac{G}{P} \cdot exp(\gamma d)$$

...3

The explained phenomena justify the sensor's response with respect to change in resistance with the applied strain. A gradual increase in resistance is seen at lower loads, followed by a steep exponential rise in the resistance at higher load ranges. The decrease in overlap area and increase in the inter-sheet distance have a majority contribution in inducing the change in the sensing layer's resistance. The above equation emphasizes the importance of the thickness and density of the sensing layer reflected in the term $G$ and $P$. Higher number of graphene sheets increases the number of parallel conductive paths causing smaller percentage changes in the parameter with applied loads. Literature shows that higher thickness requires strain input of the order of 50% to 200% for causing an appreciable change in the resistance value. A lower count of the conduction paths means a higher percentage change in the resistance with the corresponding reduction in conduction paths. However, an optimal number of conduction paths need to be established to ensure the conductive sensing layer. A visual elucidation of the above discussion can be observed in **figures 4a**, where magnified images of the GLE sensor display conductive percolative networks established between the two electrode digits at few locations on the GLE sensor. The ultra-thin layer of graphene nanosheets and the IDE pattern help visualize the conductive networks, which could not have been possible with a thicker sensing layer.

A near-linear response of the resistance change followed by exponential rise for higher strain inputs ascribes the sensing mechanism to the above two principles. A similar mechanism has been implemented by several works based on 2D nanomaterials.[51,52] The discerning factor in the peerless performance of our sensor can be explained as follows. Sensors based on nanosheets or nanomaterials warrant the minimum number of flakes to establish a conductive path between the two probing electrodes. This requirement leads to the obligation of fabricating the sensing layer as thick as necessary, which usually reaches a thickness of few microns.[53,25,48] Numerous works have proffered the use of graphene in the form of foam in composition with a binding polymer. [15,18,20–25,27,32] This approach is rather antithetical, where a 2D material is re-designed for a 3D sensor model. The natural approach of using graphene would be such that it is indeed used in a 2D model or at least near approaching 2D model. Sensors based on pristine graphene synthesized via the CVD process have also been reported. Fabrication of CVD graphene is a complex, non-scalable and expensive process and not suitable for market commercialization. Fabrication of ultra-thin sensors using 2D graphene nanosheets is deterred by its inherent random arrangement that does not form a conductive path at lower film thickness. Studies showing strain sensors using mechanically exfoliated graphene monolayer have been reported, but mass commercial-scale delivery of such sensor is un-feasible. The appropriate route to achieve ultra-thin sensors adopting graphene nanosheets is demonstrated in our work, where the randomness of the graphene nanosheets has been taken into stride, and an adaptable electrode design has been proposed. The circular IDE allows for a thinner film of graphene nanosheets where sufficient conductive paths are established between the electrodes' digits at ultra-thin film thickness. The influence of this arrangement leads to the outstanding gauge factor as predicted from the equations discussed above. A similar approach was predicted by Hempel et al., 2012,[53] where their simulation work showed that the gauge factor of a 2D material-based sensor would increase with a decrease in the material flake density. They also showed that a maximum of $>10^{18}$ sensitivity factor could be achieved via tunnelling break junction. Practical devices toward this goal have not been witnessed and would seem too ambitious, but our sensor is one of such headed in that direction.

**Applications of GLE sensor**

The GLE sensor was tested for a variety of human physiological parameters. The flexibility and compatible nature of the sensor makes it amenable to attach to the human skin and has high skin conformability due to its thin structure.

**Heart Rate:** Heart rate (HR) is one of the most vital parameters foreshowing the body's present condition. The heart's rhythmic action in pumping blood throughout the body generates pressure waves in the blood volume, particularly in the arteries. The HR can be detected by palpating the arteries that run close to the skin surface by sensing the pressure waves, aka pulses. The pulses can be probed at various locations of the body like the wrist (radial pulse), neck (carotid pulse), groin (femoral pulse) or leg (posterior tibial pulse). Convenient and comfortable measurement of pulse rate can be carried out at the radial artery (wrist) and the posterior tibial artery (ankle), as demonstrated here.

The pulse waveform recorded from the radial artery of a 33-year-old male subject is shown as in **figure 5a,5b**. The waveform indicates a resting heart rate of 66 beats per minute. The sensor was able to gather fine details of the pulse by deciphering the various phases of the heartbeat cycle. A pulse wave is generated due to the systolic wave i.e. the upstroke from the heart, and the diastolic wave, which is the pressure reflected from the limbs. The **figure 5c** shows the main systolic phase (percussion peak), the late systolic phase (tidal peak) and the diastolic wave (dicrotic peak) distinctly. The sensor showed sufficient sensitivity to detect the minute pressure pulses from the radial artery. Available PPG sensors monitor the pulse rate by volumetric measurement of oxygen in the bloodstream. The above method gives a direct quantification of the pressure exerted on the radial artery, which in turn can help in the assessment of other parameters like blood pressure. The pulse waveform shows a deflection (dV/V) of 12.5%, which is impressive and higher than biomonitoring devices reported in the literature.[54–56] The posterior tibial artery was palpated to detect the pulse rate from the ankle region **(figure 5d)**. Perceptible peaks could be recorded from the sensor measuring the HR at approximately 72 bpm **(figure 5e)**. The waveform showed a single shoulder peak significant of the dicrotic notch. Measuring pulse signals at separate locations on the body can help in determining pulse wave velocity (PWV) valuable for diagnosing arterial stiffness and other cardiovascular ailments.

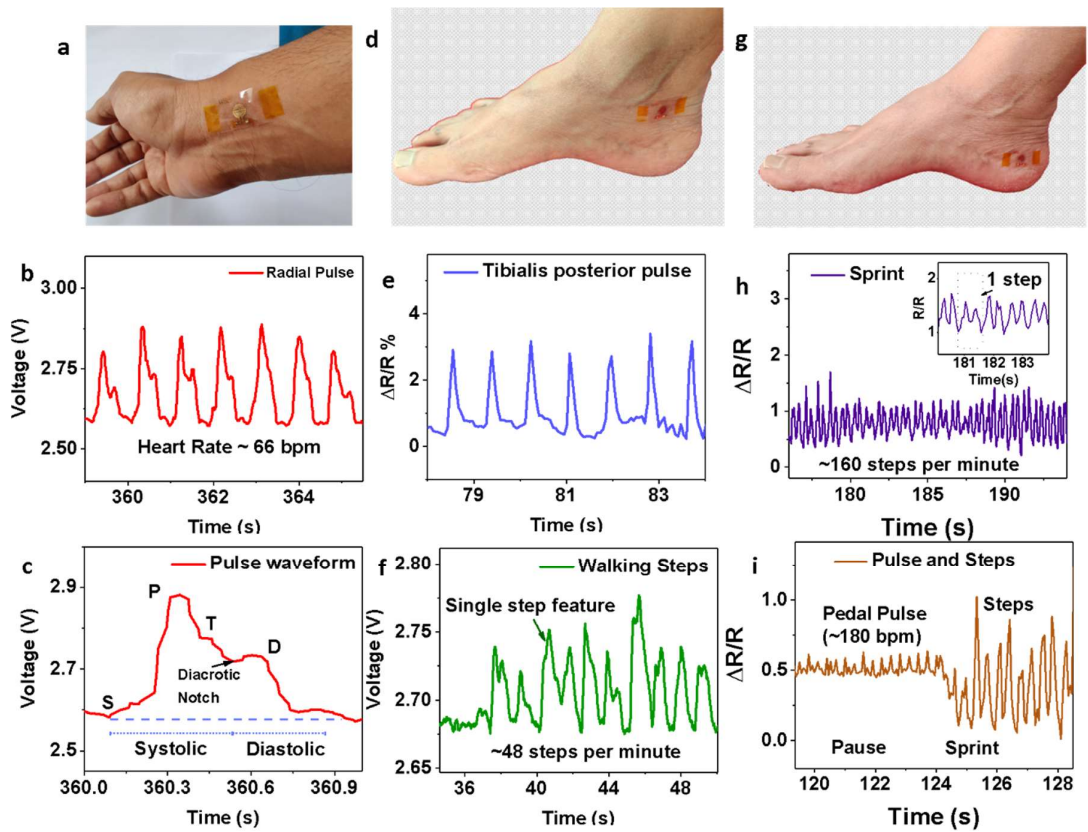

**Figure 5. (a-c)** sensor affixed on the radial artery, detection of radial pulse showing 66 bmp, expanded pulse waveform depicting characteristic features, -Start of the waveform(S),

Percussion wave(P), Tidal wave(T), Dichrotic notch, Dichrotic wave(D) and clear distinction of the systolic and the diastolic phase of the heart rhythm, (d-e) sensor affixed on the tibialis posterior artery and detection of the pulse showing only the reflected diastolic wave, (f) sensor fixed on the heel for Pedometric application, single steps are seen as local maxima, (g) detection of steps during sprinting, single step recorded as 'M' shape feature, (h) simultaneous detection of pulse rate and steps at the heel (limb extremity) during sprinting.

**Step counting:** The positioning of the sensor below the ankle on the heel also led to the identification of steps taken by the subject **(figure 5g)**. Interfacing the sensor to the data acquisition system via wireless Bluetooth communication helped in functional testing of the sensor for step counting. A peak detection algorithm was implemented in LabVIEW software to count the number of steps taken by the subject. A casual walk by the subject was recorded by the sensor as local maxima at the instance of steps taken by the subject **(figure 5f)**. The sprinting activity was detected as an 'M' shaped feature signal, probably due to complex, vigorous movements involved during this activity **(figure 5h)**. Conventional step counting wrist-worn wearable devices rely on acceleration measurement to detect swing of arms or tilt of the body, only approximating the step count. The method adopted in the reported work gives the actual step count and can be of prime importance in athletics and medicine. The positioning of the sensor at the limb extremity also led to detection of pulse rate when the subject paused during the sprinting exercise, causing the same sensor could detect pulse and steps **(figure 5i)**.

**Motion monitoring:** Electromyogram is the technique used for monitoring electrical activity pertaining to the muscles of the body. Muscle activity can also be monitored by directly detecting the mechanical force developed by the set of muscles as perceived at the skin surface. The GLE sensor was tested for routine limb movements like palm grasping **(figure 6a, 6b)** and wrist movement **(figure 6c,6d)**. Conformal tethering of the sensor to the skin surface and a broad range of strain operation led to a reliable transfer of the strain to the sensor. The sensor efficiently responded to the activities showing a high degree of sensitivity to the applied mechanical strain due to these bodily movements. The exercise leads to the sensor's possible applications toward sports, yoga and fitness training, rehabilitation of paralysis, and trauma patients.

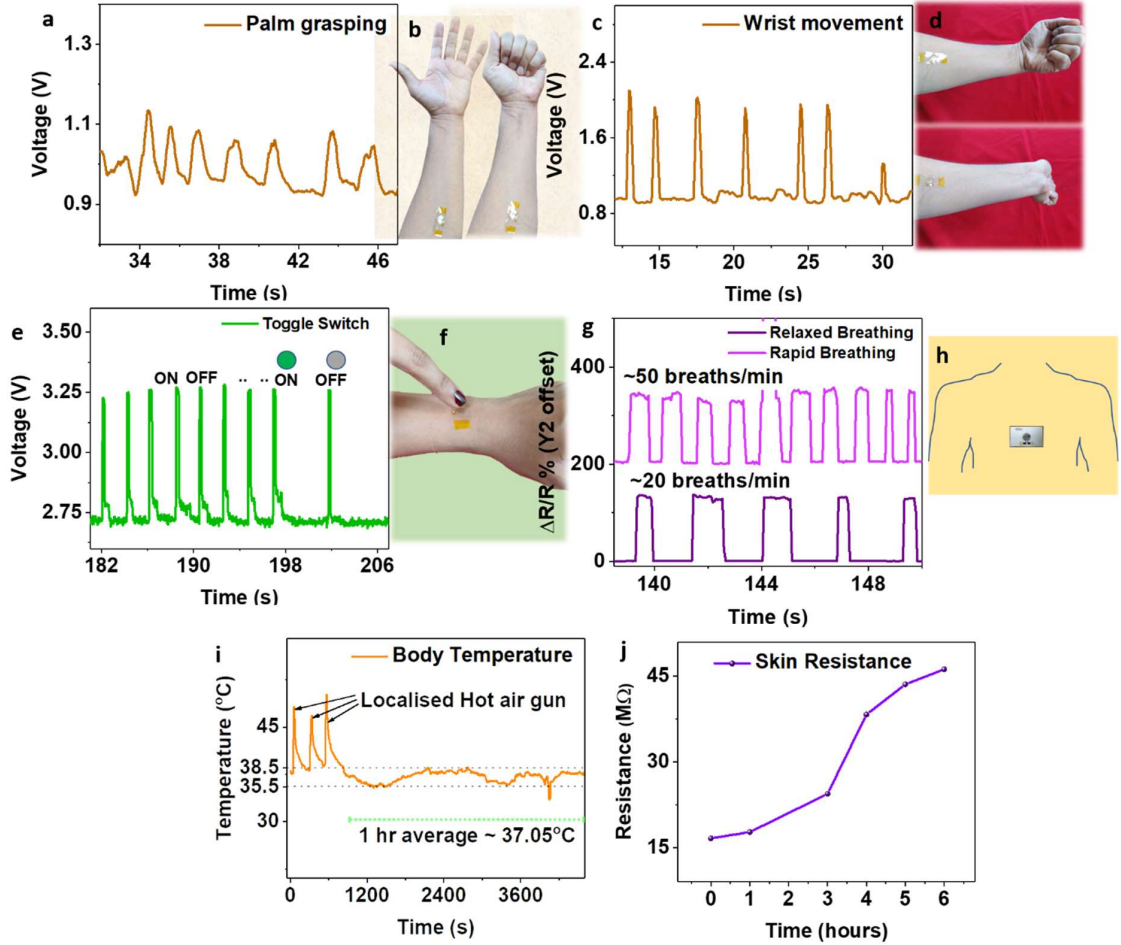

**Figure 6.** (a), (b) Detection of palm grasping movement, (c),(d) wrist movement, (e),(f) sensor as a control switch, operated by tapping on the sensor, (g),(h) measurement of breath rate showing relaxed breathing (20 breaths/min) and Rapid breathing (~50 breaths per minute) with sensor fixed on the upper abdomen, (i) body temperature measurement using GLE sensor conveying 1 hr moving average of 37.05°C and fast response to localized temperature modulations, (j) monitoring of body dehydration based on skin resistance measurement for a dehydration duration of 6 hrs.

**Sensor as a switch:** Apart from the diagnostic aspect of the wearable sensor, consumer interfacing is a favoured feature. Features like primary control of smartphone applications & IoT devices, SOS signalling require control inputs on the wearable device. The GLE sensor was tested for toggle switch control as a wearable device. The sensor was affixed on the subject's skin and asked to tap on the sensor as anyone would use a touch button. The tapping action induced a momentary change in the sensor resistance, denoted as a spike in the sensor response **(figure 6e,6f)**. The high sensitivity of the sensor allows for a feather touch input from the subject's fingertip. Repeated inputs at different forces typical of human behaviour showed corresponding identifiable response from the sensor. The signal is transmitted wirelessly to the data acquisition and processing system, i.e., LabVIEW software, where a peak detector algorithm takes these peaks as toggle inputs from the user and toggles the virtual LED indicator state (supplementary video). A similar mode of action can help in smartphone or IoT applications for ON/OFF, Play/Pause, Next, Previous, Call-receive/hang-up, and similar event inputs.

**Breathing Rate monitoring** – Breathing rate is one of the primary aspects of the respiration system. Especially in the present COVID-19 pandemic scenario, respiratory wellness has been of utmost importance. The GLE sensor was successfully tested for monitoring breathing rate by detection of the mechanical force exerted due to the rhythmic movement of the abdominal diaphragm **(figure 6g, 6h)**. The sensor was affixed near the sternum region to detect the subjects breathing pattern. The breathing rate of 20 breaths per minute and the inter-breath variability could easily be observed from the recorded waveform. The sensor showed a high ~200% change in the resistance indicating transfer of high strain due to the pronounced movement of the thoracic cavity. The subject was asked to perform a heavy physical exercise which led to heavy breathing. The recorded signal showed both time and magnitude variable features marking the breath rate of ~50 breaths per minute.

**Body Temperature Sensing:** Body temperature (BT) is one of the primary parameters crucial for the diagnosis of numerous ailments. The current COVID-19 pandemic has led to monitoring of the body temperature for each individual in every walk of life, be it a workplace, residential apartments or a group gathering setup. Every individual needs to look out for symptoms like fever, even while at home. Current body temperature investigating devices are one-time body contact type devices like mercury thermometers, thermistor temperature meters etc. Temperatures are taken from openings on the body, such as the mouth, rectum, or ear. Readings are taken from skin surfaces where the ambient air interaction is lower, such as the armpit or groin. Non-contact type devices like IR thermometers provide rough temperature measurements by directing toward areas with low-fat insulation like the forehead or areas with arteries close to the skin surface like inside of the wrist. While the accuracy of the measurement depends on the type of meter and area of probing, the task takes some time and not suitable for continuous recording of the parameter. A continuous body temperature profile often provides crucial information in diagnosing diseases and prescribing the right treatment. Very few wearable body temperature devices are commercially available because of the complexity in measuring skin temperature, which is often influenced by the surrounding environment. The currently available metal and semiconductor type thermometers are bulky and lack skin confirmability. These devices are adversely affected by the surrounding temperature since they are 3D structured with only a single side in contact with the skin surface. A reliable and fast body temperature measurement can be achieved using 2D materials like graphene. The GLE sensor has been inspected for body temperature measurement and found to be highly effective. The GLE sensor for temperature measurement was fabricated on a PET substrate ($\rho$ = 1390 kg/m$^3$, thermal conductivity@23°C = 0.14 W/m.K, Specific heat capacity ($c_p$) = 1.27 kJ/kg.K, tensile strength = 70 MPa) and covered with a PDMS coating layer ($\rho$ = 970 kg/m$^3$, thermal conductivity@23°C = 0.15 W/m.K, Specific heat capacity ($c_p$) = 1.46 kJ/kg.K, tensile strength = 2.24 MPa). The higher strength of the PET substrate minimized the strain-dependent response of the sensor as compared to PDMS substrate GLE. The PET substrate use ensured that the small mechanical deformation of the sensor did not affect the sensor response. The use of PDMS for temperature sensor was deemed unfit since small un-intended deformations induced in the sensor resulted in a considerable change in the response. The linearized equation from the temperature response plot in figure 3f was programmed into the data acquisition, and real-time body temperature was measured and recorded in **figure 6i**. The sensor responded to an external temperature stimulus of a hot air gun, where the local temperature momentarily increased to ~50°C and equally felt by the subject. The temperature quickly returned to normalcy which was tracked by the sensor. The sensor recorded the minute changes in the body temperature during a one-hour duration, with the average temperature recorded at 37.05°C. The 2D nature of the graphene nanosheets provides low thermal inertia, providing high sensitivity and improved body temperature tracking. A barrier of thin PDMS covering helps protect the device from coming in direct contact with the skin. A thicker insulation barrier of PDMS (200um) between the graphene nanosheets and the surrounding environment minimized heat transfer from the environment to the sensor, while the thinner PET substrate (50um) enabled faster heat conduction from the skin surface to the graphene nanosheets.

**De-hydration detector:** Skin impedance is a potent measure of estimating the body dehydration level. The presence of water in the body and essential electrolytes make the skin and underlying epidermis conductive. The percentage of water content in the skin modulates its conductivity and is a good measure of the body hydration level. The Laser-etch process was conveniently used to pattern two electrodes for skin contact. The resistance of the epidermis of a 33-year-old subject was measured half an hour after hydration (drinking half a litre of water). The skin resistance was measured consequently after each hour for a duration of 6 hours, and the subject did not drink water for the above duration. The measurements reflect an increase in skin resistance due to dehydration **(figure 6j)**. The procedure carried out emulates a standard body impedance measurement setup. However, the skin contact impedance of the gold pads could not be minimized since the conductive electrolytic gel was avoided. The use of conductive gel was deemed impractical for a day-to-day wearable device. Further study in the development of regular use wearable dehydration sensor can be explored based on the demonstrated work.

**Wearable fashion accessory:** Incorporation of multiple sensors in a wearable device is a complex task. The sensor size, shape, and placement depend on various factors and limitations in realizing a multifunctional wearable device. The inclusion of multiple sensors in a wearable device makes it bulky and unyielding. The literature on sensors based on nanomaterials shows sensors which are bulky and look like patch (dark or transparent) on the skin surface. The devices are un-aesthetic, thick, lack scalability and not suitable for designing a marketable wearable device. Using these sensors for multiple parameters can lead to further clattering and a shabby appearance. The reported procedure of fabricating the GLE sensor provides an excellent opportunity to design a multifunctional, mass scalable, versatile, and fashionable wearable device. The GLE sensor has a nearly transparent, imperceptible graphene layer that allows the gold electrodes to be seen clearly through the PDMS/PET substrate and coating. The gold electrodes can be conveniently designed as per the required sensor specifications with an eye for the art. A wearable bracelet has been designed and developed using the above procedure **(figure 7a-c)**, consolidating multiple sensors for varied applications. The array of sensors is a facile integration of sensors without compromising the simplicity of the device fabrication technique while improving the device aesthetics. The bracelet bears the body temperature sensor, dehydration detector, pulse rate sensor on the underside of the wrist. Two sensors on the lateral side of the wrist provide for detecting wrist movement. Three GLE sensors are provided on the upper wrist area as control switches and wrist movement detectors. The bracelet is fabricated on a sandwich substrate of PET and PDMS, where the PET includes through holes in the area of GLE sensor, allowing adequate strain transfer to the sensor. The temperature sensor and the dehydration detector are placed on the PET substrate. A human-machine interface (HMI) screen developed in LabVIEW software shown in figure 7e, displays real-time data from all the sensors and processed information to indicate the current body temperature, heart rate, state of hydration and inputs from the user. The GLE band showed cross-sensitivity for sensors dedicated to the control buttons and wrist movement. Post processing of these signals using machine learning algorithms can help in resolving this issue and further assist in recognizing movement patterns.

A similar specimen has been fabricated as a wearable ankle device with three devices for pulse rate, step counting and body temperature sensing each **(figure 7d)**. The device can be incorporated into smart, trendy sports shoes with vital biomonitoring capabilities.

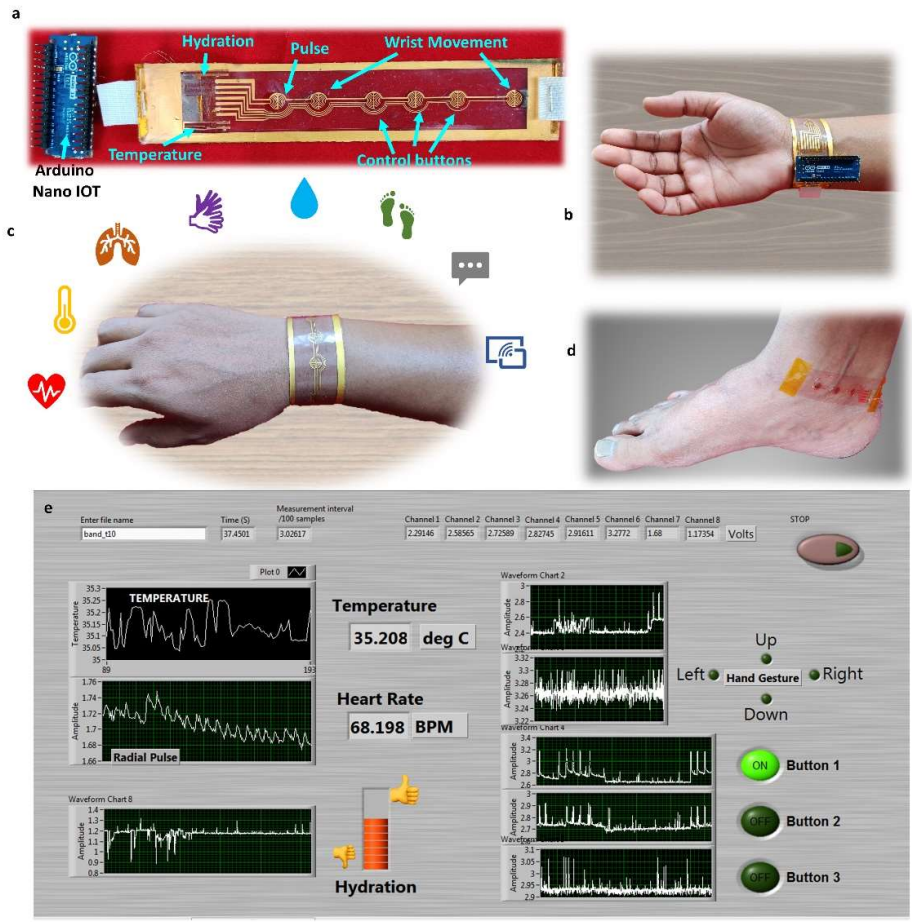

**Figure 7.** (a) GLE Band with integrated sensors for detecting HR, BT, Hydration, wrist movement and tactile inputs, interfaced with Arduino nano 33 IOT, (b),(c) GLE Band worn on hand, icons highlighting applications of GLE sensor (from left to right) monitoring of heart rate, body temperature, breath rate, limb movements, dehydration, steps, relaying control commands and sensor data over wireless communication, (d) multifunctional device for foot

based HR, BT and steps for smart shoes, (e) Customized HMI screen simultaneously displaying vital parameters and tactile inputs from the integrated GLE band,

Reports on nanomaterial-based wearable sensors have seldom considered the aesthetics, marketability, and commercial prospects of the device. The regions outside the sensor area of the GLE band can be customized with artworks and personalized designs and messages, as shown in the Egyptian cuff styled GLE bracelet depicted in **figure 8a, 8b**. Keeping the basic structure of electrodes intact, the gold thin film can be fashioned and customized as per consumers palate. Designs like personalized tattoo designs, memorabilia symbols, company logo/advertisements, ID-card, person identification and details, especially for the aged folks, bar-coded information and so forth **(figure S)**, can be adapted to meet the objective. The device is further customizable by employing thin films of different metals like platinum, copper and other affordable conductive metals and alloys. Wearable sensors, even with acceptable performance characteristics, can not be preferred as a daily wear consumer product unless they provide charm and comfort to the beholder. The presented device blends the superior performance of GLE sensors with a fashionable design attracting appreciable mass appeal, especially to the womenfolk.

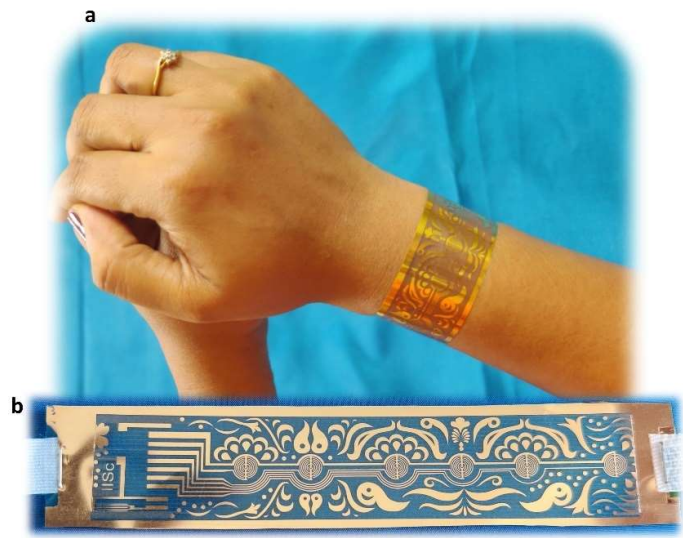

**Figure 8.** (a), (b) Customized gold GLE Bracelet as a Smart Biomonitoring trendy gadget.

**Table 1:** Comparison of few wearable devices reported in literature with the currently presented work.

| Material | Method of fabrication | Sensitivity/GF | Sensing layer Thickness | Appearance | Multifunctional | reference |
|---|---|---|---|---|---|---|
| **Graphene nanosheets** | Drop cast on cicular IDE | 6.38E+07 | ~10nm (upto 50 nm) | Transparent layer, Gold pattern | Yes | This work |
| **Hollow tubing graphene fibers** | CVD graphene with PDMS | 48.9 | | Dark strand | No | [7] |
| **Graphene Aerogel** | Vacuum Freeze drying | 159 | 1 cm3 | Dark Sponge | No | [22] |
| **Graphene membrane** | Air drying | 52.3 kPa-1 | 10 um | Dark Patch | No | [25] |
| **rGO** | Ecoflex filled with rGO ink | 31.6 | | Dark channel in rubber | No | [27] |
| **Gold nanoparticles** | Gold sputter coating on PDMS | 5888.89 | 50 nm | Gold layer | No | [37] |
| **PEDOT:PSS/Xyl on ESPVPANI** | Spread coating | 26 | 50 um | Dark nanowire mat | Yes | [40] |
| **rGO scales** | Spray coating | 1054 | 300 um | Dark strip | No | [47] |
| **AgNWs/rGO on TPU** | Spray coating | 4.40E+07 | 74 um | Dark strip | No | [48] |
| **Graphene nano platelets** | Dip coating silicone rubber | 10 | Bulk ->1cm | Dark Sponge | No | [55] |
| **Braided graphene belts** | CVD graphene | >175 | 525 nm | Dark patch | No | [57] |
| **AgNW/WPU-MXene** | PU fibre dipcoating | 1.60E+07 | | Dark strand | No | [58] |
| **Au/Cu/ZnO/Pt Thin-film** | Clean room techniques | 18 | ~100 um | Transparent gold pattern | Yes | [59] |
| **rGO/TPU** | Ultrasonication | 79 | 200um | Dark Strip | No | [60] |
| **CNT** | Blade coating | 1140 | 7.5um /300um | Transparent | No | [61] |
| **Nanographene** | PECVD | 507 | | Transparent with glod pads | No | [62] |

The **table 1** reveals a superlative performance of the reported work here with respect to the ease of fabrication, sensitivity, sensing layer thickness, multi-functionality and the appearance aspect. The device has been demonstrated with the added feature of wireless communication of the sensor data. The acquired data has been processed runtime in LABVIEW but can be comfortably implemented in smartphones and other smart devices.

**Discussion**

The graphene nanosheets on laser-patterned IDE helps in realizing ultra-thin sensing layer via a facile cost-effective approach. The innovative design resulted in ultra-high record sensitivity with repeatable outcomes over different experiments. The sensor had an excellent low-strain resolution, better than 0.02%, effective in detecting minute pressure signals. The GLE sensor was tested for multifunctional roles of strain and temperature sensing. Biomonitoring applications like measurement of heart rate, breathing rate, body temperature, dehydration, steps and hand movements have been successfully exhibited to establish the immense potential of the sensor. The laser-patterning method has been shown advantageous in fabricating an assemblage of sensors for multifunctional wearable devices. The device is interfaced wirelessly with a data processing system to gather crucial information, an increasing necessity considering the ongoing COVID-19 pandemic. A multifunctional, comfortable, routine-use market-ready smart, fashionable bracelet has been fabricated. The standout feature of the GLE sensor is reported to be ultra-high sensitivity, low-limit strain resolution, thin, skin-conformal structure, biocompatible and an aesthetic appeal posing as a genuine marketable fashion accessory.

**Experimental section**

**Material Synthesis:** Synthesis of Graphene oxide was carried out via the modified Hummer's method.[63] GO was thermally exfoliated using microwave treatment in an LG Microwave oven at maximum power (800W) for 1 minute.[64] The GO instantly heats up and explodes with bright illumination. A considerable increase in the volume of the MEGO material is observed compared to the precursor GO. The MEGO particles are lighter and darker in appearance. The Graphene nanosheet ink is prepared by dispersing MEGO in NMP in a pre-defined ratio followed by 30 minutes of ultrasonication. The heavier particles in the solution settle at the bottom while lighter graphene nanosheets form a uniform ink with the solvent. This translucent ink is separated and used to drop-cast on the sensor pattern.

**Sensor fabrication:**

Sensors with different designs and dimensions have been attempted using the Laser-etch process. The two types of substrates employed for sensor fabrication are PDMS and PET. The PDMS polymer was prepared from the combination of base material and curing agent and coated on a glass slide/ PMMA platform. The thickness of the PDMS layer is controlled using a spin-coater to obtain ~60 µm thick layer @1000rpm. The PDMS is heat-treated at 90°C for 4 hrs allowing it to dry. The PDMS layer is then peeled from the slide and placed back on the slide in order to release its internal stress and convenient to peel the sensor later on. PET substrate of 50 µm thick is used specifically for temperature and dehydration detector. A thin-film stack of Chromium (10nm) and Gold (125nm) is deposited on the prepared PDMS or PET substrate without breaking the chamber vacuum in a dual target RF/DC Magnetron sputtering system. Laser-trimming of the deposited thin-film is achieved using a fibre laser source (Light mechanics LM-BT-20, 1060 nm wavelength) at optimized power parameters such that the thin-film is etched, but the substrate remains intact. The Laser-trimming pattern is generated using Autocad software, which provides the flexibility for easy sensor design. A circular IDE pattern has been chosen as the strain sensor electrode design, which allows direction independent strain sensitivity. The IDE arms are ~200um thick with an equal amount of gap between the electrode digits. The surface of the sensor is plasma cleaned for 2 mins to obtain a hydrophilic surface. The GLE sensor ink is drop-casted on the sensor and heat-treated at 80°C for 4 hrs to dry. The nanosheets randomly settle on the IDE and form multiple conductive paths. Leads are taken using copper wires and conductive silver epoxy. The sensor is then coated with another layer of PDMS as protection from external factors. The temperature sensor is fabricated using the PET substrate and conventional IDE electrodes. The impedance sensor is a set of two electrodes without the graphene layer.

**Mechanical and Thermal characterization:** Mechanical experiments on the GLE sensor were performed using the combination of Newport motion controller and motorized linear stage (EPS 301 and M-ILS150CC respectively) controlled via MATLAB programming. The GLE sensor was fixed vertically i.e. short side edges perpendicular to the stage. A rounded probe was then used to induce known deflections at the centre of the GLE sensor, right at the location of the IDE.

**Experimental Setup:** The electrical measurements are taken using Keithley 6517B High-resistance meter for resistance measurement and Arduino Nano 33 IOT when the sensor resistance is measured in the form of voltage drop. The Arduino provides wireless transmission of the measured parameters facilitating sensor mobility and demonstration of prototype wearable biomedical device. The data acquisition is performed via Labview with programming capabilities for runtime analysis and recording step count, Control switch operation and body temperature.

**Supporting Information**

Additional information depicting figures supporting Material synthesis, Device Fabrication, experimental setup, and methods are given in supplementary information. Additional information regarding AFM, XRD, Optical profilometry characterizations is furnished. Additional graphs supporting the results discussed in manuscripts are available in Electronic Supporting Information.

**Author Contributions**

VS conceptualized, designed and fabricated the devices. Experiments, characterizations and technical discussions were carried out with the assistance of SN, VK and SK. All authors contributed in preparing the manuscript. The final version of the manuscript has been approved by all the authors. The authors declare that they have no competing financial interests.

**Author Information**

Corresponding Author: **Konandur Rajanna**

Department of Instrumentation and Applied Physics, Indian Institute of Science, Bangalore 560012, India.

**Acknowledgment**

The authors thank the Micro-nano characterization facility (MNCF) and Advanced Packaging Lab at Center for Nano Science Engineering (CeNSE), IISc for their characterization facilities and device fabrication facility respectively. The authors are grateful to the assistance provided by Mr Veera Pandi N, from Advanced Packaging lab, CeNSE, IISc and Mr Manu Pai from Dept of Instrumentation and Applied Physics, IISc. We would like to thank Mrs Snehal Katti for her contribution on the art-work designed on the bracelet.